\documentclass[prl,aps,twocolumn,floats]{revtex4}
\usepackage{amsmath}
\usepackage{graphicx}
\usepackage{epsfig}

\newcommand{\be}{\begin{equation}}
\newcommand{\ee}{\end{equation}}
\newcommand{\bea}{\begin{eqnarray}}
\newcommand{\eea}{\end{eqnarray}}

\newcommand{\comma}{,}
\newcommand{\gm}{}

\begin{document}
\title{Dissipative quantum phase transition in a quantum dot}
\author{ L\'aszl\'o Borda$^{1}$, Gergely Zar\'and$^{1,2}$, and D. Goldhaber-Gordon$^3$}
\address{
$^1$ Department of Theoretical Physics
and Research Group ``Theory of Condensed Matter'' of the Hungarian Academy of Sciences\comma Budapest University of Technology and Economics\comma
Budafoki \'ut 8.  H-1521 Hungary \\
$^2$  Institut f\"ur Theoretische Festk\"orperphysik\comma
Universit\"at Karlsruhe\comma  76128 Karlsruhe, Germany  \\
$^3$ Physics Department and Geballe Laboratory for Advanced
Materials\comma Stanford University\comma Stanford CA 94305\comma
USA}
\date{\today}

\begin{abstract}
We study the transport properties of a quantum dot
(QD) with highly resistive gate electrodes, and show that the QD
displays a quantum phase transition analogous to the famous
dissipative phase transition 
first identified
by S. Chakravarty [Phys. Rev. Lett. {\bf 49}, 681-684 (1982)]; for a
review see
[A. J. Leggett {\em
et al.}, Rev. Mod. Phys. {\bf 59}, 1 (1987)]. At temperature $T=0$,
the charge on the central island of a conventional QD changes
smoothly as a function of gate voltage, due to quantum fluctuations.
However, for sufficiently large gate resistance charge fluctuations
on the island can freeze out even at the degeneracy point, causing
the charge on the island to change in sharp steps as a function of
gate voltage. For $R_g<R_C$ the steps remain smeared out by quantum
fluctuations. The Coulomb blockade peaks in conductance display
anomalous scaling at intermediate temperatures,
and at very low temperatures a sharp step develops in the QD conductance.
\end{abstract}


\maketitle

The single electron transistor (SET) is one of the most basic
mesoscopic devices: A conducting island or quantum dot is attached by tunnel barriers 
to two leads and a
capacitively-coupled gate electrode sets the number of electrons on
the dot.  {\gm For low enough temperatures, $T\ll E_C$, charge fluctuations
of the dot are suppressed except when the gate is tuned to make two
charge states nearly degenerate. At
 these ``charge degeneracy points'' the charge on the dot
strongly fluctuates.  For typical metallic SETs with a very large
number of tunneling modes  quantum fluctuations of the charge turn
out to be suppressed at low temperatures \cite{Metal_SET_reviews,Metal_SET,Metal_SET_experiments}.
For semiconducting SETs with {\em single mode junctions}, however,
quantum fluctuations of the charge are important and broaden out the
charging steps at low $T$: In the limit of vanishing level spacing,
$\delta \epsilon\to 0$ charge fluctuations are described by the
two-channel Kondo model
 \cite{Matveev},  while for $T\ll  \delta \epsilon $ one recovers the so-called
``mixed valence'' regime of the Anderson model \cite{Costi}.}

{\gm In the above discussion we neglected the effect of Ohmic
dissipation  in the lead electrodes. While this has been extensively
studied for SETs with a very large number of tunneling modes
\cite{Metal_SET}, there is much less known about the effects of
dissipation in the Kondo regime: In  a recent paper Le Hur showed
that, assuming a continuum of quantum levels on the SET and a single
tunnel mode,  coupling to a dissipative bath drastically modifies
the results of Ref.~\cite{Matveev}: large enough dissipation drives
a Kosterlitz-Thouless-type {\em phase transition} and leads to a
complete suppression of charge fluctuations even at the degeneracy
point~\cite{LeHur,Pascal}. However, for most semiconducting devices
the level spacing $\delta\epsilon$ {\em cannot} be neglected in
comparison to temperature, and spin fluctuations must also be
considered. Here we shall therefore investigate the effects of
dissipation at temperatures {\em far below} the level spacing on the
dot, $T\ll \delta\epsilon$, a more realistic low-temperature limit
for typical semiconductor SETs. As we show below, a
dissipation-induced quantum phase transition takes place for $T \ll
\delta\epsilon$ as well, although with different and more
complicated properties due to the interplay of charge and spin
fluctuations,  and at a larger dissipation strength (gate resistance)  
than that
needed for $\delta\epsilon \to 0$ \cite{LeHur,Pascal, Karyn}.}

In this $T\ll \delta \epsilon$ regime, the coupling of the quantum
dot to the gate voltage is usually described by the Hamiltonian, \be
H_{\rm dot} = E_C\; \Bigl(\sum_\sigma d^\dagger_\sigma d_\sigma -
n_g\Bigr)^2\;, \label{eq:H_dot} \ee where $E_C \equiv e^2/2C_\Sigma$
denotes the charging energy, with $C_\Sigma$ the total capacitance
of the dot, and $e$ the electron charge. We retain only one
single-particle level $d$, and we assume that it is empty or singly
occupied, depending on the dimensionless gate voltage,
$n_g$~\cite{footnote1}. Assuming
weak coupling between the dot and the source and drain electrodes,
charge transfer can be described within the tunneling approximation,
\be H_{\rm tun} = V \sum_\sigma \int d\epsilon\; \Bigl(\sum_\sigma
d^\dagger_\sigma \psi_\sigma(\epsilon) +{\rm h.c.}\Bigr)\;,
\label{eq:T} \ee were $\psi_\sigma(\epsilon)$ annihilates an
appropriate linear combination of left and right lead electrons of
energy $\epsilon$ that hybridize with the dot state $d_\sigma$, and
satisfies the anticommutation relation
$\{\psi_\sigma(\epsilon),\psi_{\sigma'}(\epsilon')\} =
\delta_{\sigma\sigma'}\delta(\epsilon-\epsilon')$ \cite{footnote2}.
Throughout this paper we assume
that the quantum dot is close to symmetrical but our analysis
carries over easily to asymmetrical dots as well.

Eqs.(\ref{eq:H_dot}) and (\ref{eq:T}) are thought to provide a
satisfactory description of the SET  for $T \ll \delta
\epsilon$ for most experimental situations studied so far,
including the Kondo regime \cite{Costi}. However,
Eq.~(\ref{eq:H_dot}) does not account for the {\em relaxation of
electrostatic charges} in the nearby electrodes: in reality, when an
electron tunnels into the dot, an electrostatic charge $\delta Q = e
C_g/C_\Sigma$ is also generated on the gate. Transferring this
charge from the outside world to the gate electrode 
through a shunt resistor requires time,
and creates dissipation \cite{footnote3}.
Consequently, tunneling between dot and leads will be
suppressed by Anderson's orthogonality catastrophe. The simplest way
to account for this shunt resistance  is to add a term 
\cite{Metal_SET_reviews} 
\be H_{\rm
diss} = \lambda
\;\bigl(\hat n - \frac {2}{3}\bigr) \;\varphi\;, \label{eq:diss}
\ee
where the operator $\hat n \equiv \sum_\sigma d^\dagger_\sigma
d_\sigma$ measures the number of electrons on the dot, and the
bosonic field $\varphi$ describes charge density
excitations in the gate electrode participating in the screening
process, with their imaginary time correlation function defined as
$\langle T_\tau \varphi(\tau) \varphi(0)\rangle = 1/\tau^2$ \cite{footnote4}.
The dissipation strength $\alpha \equiv \lambda^2/2$ can be
estimated following the procedure of Ref.~\cite{PofE}, and we
find
\be
\alpha = \frac {C_g^2 }{4\;C_\Sigma^2} \frac{R_g} {R_Q}\;,
\ee
with $R_g$ the low frequency resistance of the  gate electrode,
and $R_Q = h/2e^2$ the resistance quantum. In the absence of
dissipation, Eqs.(\ref{eq:H_dot}) and (\ref{eq:T}) would predict
that the average charge on the island changes smoothly from $\langle
\hat n\rangle\approx 0$ to $\langle \hat n\rangle\approx 1$ around
$n_g \approx 1/2$ due to the presence of strong quantum fluctuations
at $n_g \approx 1/2$ \cite{Costi}. As we shall see below, the
properties of this transition may be dramatically modified for $\alpha
>1/2$, producing a {\em sharp step} in $\langle \hat n\rangle$ at $T=0$.

Since we shall focus on the vicinity of the charging step,
$n_g\approx 1/2$,  we  restrict the Hilbert space to the empty
state $|0\rangle$ and the two singly-occupied states $|\sigma
\rangle \equiv d^\dagger_\sigma |0\rangle$ of the dot, and
approximate the sum of Eqs.~(\ref{eq:H_dot}), (\ref{eq:T}) and (\ref{eq:diss}) as
\bea
\tilde H =  V\sum_\sigma \bigl( |\sigma \rangle\langle0| \psi_\sigma
+ {\rm h.c.}\bigr) - \Delta\; \hat Q + \lambda \;\hat Q \;
\varphi\;, \label{H_sum}
\eea
where $\Delta \equiv E_C(1-2n_g)$
measures the difference between
{\gm the classical energy of the two charge states of the dot},
the operator $\hat Q \equiv
( |\uparrow \rangle \langle \uparrow| + |\downarrow \rangle \langle
\downarrow|
 - 2 |0 \rangle \langle 0 | )/3$ measures the difference between the charge on the dot and
its value at the charge degeneracy point, and $ \psi_\sigma = \int d\epsilon\; \psi_\sigma(\epsilon)$.

To map out the complete phase diagram of the model, we bosonize the
above Hamiltonian \cite{Delft},  and perform a renormalization group
analysis using an operator product expansion method \cite{Cardy}.
Although we use different methods, the derivation of the scaling
equations is similar to that in Ref.~\cite{Kotliar}. In fact, a
mapping between the present model and the one studied in
Ref.~\cite{Kotliar} enables us to carry over much of the analysis as
well. Intriguingly, a new exchange coupling $j$  is also generated
under scaling\cite{Kotliar}:
\be
H_K = \frac j2 \sum_{i,\alpha,\beta}\; S^i
\;(\psi^\dagger_\alpha \sigma^i_{\alpha\beta} \psi_\beta)\;,
\ee
where the spin operators denote $S^i = \frac 12 \sum_{\alpha,\beta}
|\alpha\rangle \sigma^i_{\alpha\beta} \langle \beta|$, with
$\sigma^i$ the Pauli matrices. To lowest order, the scaling equations
read \cite{footnote5}
\bea \frac{dv}{dl} & = & \bigl(\frac 12
-\alpha)\;v + \frac{3j}4 \; v + \dots\;, \label{eq:v}
\\
\frac{dj}{dl} & = & j^2 + 2\; v^2 + \dots\;,
\label{eq:j}
\\
\frac{d\alpha }{dl} & = & -3 \bigl(\frac 12 + \alpha)\;v^2
+ \dots\;,
\label{eq:alpha}
\\
\frac{d\tilde \Delta }{dl} & = &
\tilde \Delta - \frac 38 j^2 + v^2 + \dots\;,
\label{eq:delta}
\eea
where $l = \ln\;a$ is the scaling variable, and
 $v = V a^{1/2}$ is the dimensionless tunneling rate
with $a$ a microscopic time scale initially of the order of $a_0
\sim 1/\delta\epsilon$. The coupling $\tilde \Delta \equiv a \Delta$
is the dimensionless splitting of the charge states, slightly
renormalized by the dissipative coupling.  Note that these equations
are exact in the dissipative coupling $\alpha$, but they contain
contributions in the small couplings $v$ and $j$ only up to second
order.

\begin{figure}[tb]
\begin{center}
\includegraphics[clip,width=8cm]{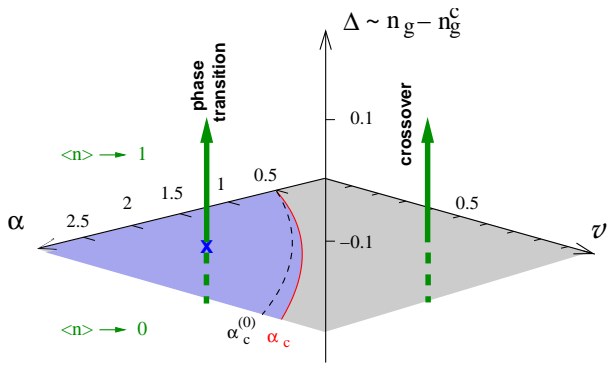}
\end{center}
\vskip0.1cm
\caption{\label{fig_phase} Schematic phase diagram of the SET in
the presence of dissipative coupling. $\alpha_c$ denotes the critical value of
$\alpha$, while  $\alpha_c^{(0)}$ is its value obtained by neglecting
the generated exchange coupling $j$. For $\alpha > \alpha_c$ there is
a phase transition from $\langle n\rangle=0$ to $\langle n\rangle=1$,
while in the more familiar situation of weak dissipation there is
a crossover.
}
\end{figure}

Following Ref.~\cite{Kotliar}, we first neglect the effect of the
exchange coupling $j$ and focus on the vicinity of the degeneracy
point, defined through the condition $\tilde \Delta(l\to\infty) =
0$.
As we shall see, this approximation describes  much of the
experimentally-relevant regions, because the Kondo temperature
associated with spin fluctuations turns out to be usually quite small in the
vicinity of the phase transition, $\alpha \approx \alpha_c$, even
for single occupancy. Within this approximation we can solve the
scaling equations analytically \cite{unpub}, and for large enough
$\alpha$ we find that as $T\to0$ the effective tunneling $v$ scales
to zero, while $\alpha$ scales to a finite value $\alpha_\infty$
larger than 1/2. Thus, quantum fluctuations of the dot charge {\em
vanish} as $T\to0$. In the above approximation, the relation between
the corresponding critical value of the parameter $\alpha =
\alpha^{(0)}_c$ and the level width $\Gamma \approx 2\pi v^2
\delta\epsilon$ is determined by the equation \be \frac {\Gamma}
{2\pi\;\delta\epsilon}\approx v^2 = \frac 23\Bigl( \alpha^{(0)}_c -
\frac 12 - \ln \bigl( \alpha^{(0)}_c + \frac12\bigr)\Bigr)\;, \ee
and is  plotted in Fig.~\ref{fig_phase}. For small values of $v$ we
find $\alpha^{(0)}_c(v) = 1/2 + \sqrt{3} v + v^2 + \dots$, and the
transition is of Kosterlitz-Thouless type \cite{Kosterlitz}: On the
localized side, $\alpha> \alpha^{(0)}_c$ (or $\Gamma<
\Gamma^{(0)}_c$), at the degeneracy point the height $\delta G$ of
Coulomb blockade peaks scales to zero as a power law \cite{unpub},
\be {\delta G(T) \over G_Q} \sim \; v^2 \sim \bigl(\frac T
{\delta\epsilon}\bigr)^{2\alpha_\infty-1} \ee with $G_Q$ the quantum
conductance. On the metallic side, on the other hand, quantum
fluctuations always dominate and preserve conductance even at $T=0$,
though near the transition the conductance shows a non-monotonic
behavior: $\delta G(T)$ first slowly decays and then starts to
increase below a temperature $T^*$ that vanishes exponentially as
one approaches the phase transition, $
 T^* \approx \delta\epsilon\; \exp\{ -\pi/2({\alpha_c^{(0)}}^2 -
\alpha^2)^{1/2}\}
$, until  finally a mixed valence state with a
large conductance is formed at a temperature $T^{**}\sim
{T^*}^2/\delta\epsilon \ll  T^*$. For the critical value of
$\alpha$, $\delta G$  decays to zero
logarithmically,
\be
\delta G(T,\alpha = \alpha_c^{(0)}) \sim  G_Q \;
\frac 1{\ln^2(\delta\epsilon /T)}\;.
\ee

The gate-voltage dependence of the expectation value of
the charge on the dot,  $\langle \hat Q\rangle$,  can be directly
detected by another SET or point contact electrostatically coupled
to the dot.
So far, we only considered the degeneracy point,  $\tilde \Delta (\infty) = 0$.
However, in the localized phase, $\Delta $ is a {\em relevant perturbation} and
generates a first order phase transition at $T=0$ (see Fig.~\ref{fig_phase}). Correspondingly,
$\langle \hat Q\rangle$ {\em jumps} \cite{Cardy} at the degeneracy point. At
finite temperatures the width $\delta n_g$ of this charge step becomes
finite, $\delta n_g\sim T/E_C$.

On the delocalized side $v$ scales to large values and $\alpha\to 0$
so that the dissipation effectively disappears from the problem,
and we recover  the well-understood mixed valence fixed point of the Anderson
model~\cite{Costi,Wilson}. At this fixed point the charge susceptibility is
finite,  and thus the charging step remains smeared out by quantum
fluctuations even at $T=0$. The width of the step is roughly
determined by  the scale $T^{**}$, which vanishes exponentially fast
as we approach the transition. The corresponding $T=0$ phase diagram
is sketched in Fig.~\ref{fig_phase}.

We shall now discuss how the thus-far neglected exchange coupling
$j$ changes the above picture.  Let us again first examine the
degeneracy point, $\tilde \Delta \to 0$, and let us assume that we
are deep in the localized phase. Here charge fluctuations ($v$) are
irrelevant, however  a  small exchange coupling is still generated
at the beginning of scaling, and produces a spin Kondo effect at the
degeneracy point at some  Kondo temperature $T_K$. {\gm The  scale
$T_K$ is typically small, but could likely be pushed into the
measurable range by gradually opening up the quantum dot.} Although
the exchange coupling somewhat renormalizes $v$, it will typically
not make it relevant. In other words, in the localized phase, at the
degeneracy point one scales at low $T$ to a special quantum state
with large spin fluctuations but suppressed charge fluctuations
\cite{Kotliar,Kotliar2}. As a consequence, at the degeneracy point
the conductance of the SET becomes  of the order of $G_Q$ at
temperatures $T\ll T_K$, even though charge fluctuations of the SET
are suppressed and $v$ scales to zero. Despite the strong spin
fluctuations, $\Delta$ remains a relevant perturbation, and
therefore, at $T=0$ there is still a {\em jump} in $\langle \hat
n\rangle(n_g)$. Interestingly, on the $\langle \hat n\rangle\approx
1$ side of this jump, the Kondo effect always develops, the spin is
screened, and a Fermi liquid is formed for $T\ll T_K$
\cite{Nozieres}.  Therefore the $T=0$ conductance is close to the
quantum conductance. On the other side of the transition,
however,$\langle \hat n\rangle\approx 0$, spin fluctuations are
suppressed, and no Kondo effect takes place. Since the conductance
is directly related to the scattering phase shifts and thus the
occupation number $\langle \hat n\rangle$ through the Friedel sum
rule, the $T=0$ conductance in this localized phase must also have a
{\em jump} as a function of $n_g$. This regime may be difficult
to explore experimentally since we expect $T_K$ to be small unless
both $\alpha$ and the tunnel coupling $V$ are sufficiently large.

Exchange fluctuations play another important role too: they increase
the value of $v$ and,  as a consequence,  the true critical value of
$\alpha$ is somewhat renormalized,  $\alpha_c^{(0)}\to \alpha_c$. It
is very hard to estimate this critical value of $\alpha_c$, but for
vanishingly small values of $v$ we have been able to show that
$\alpha_c(v\to0) = \alpha^{(0)}_c(v\to0)=1/2$.
For non-vanishing values of $v$, counter-intuitively, $\alpha_c$ seems to be shifted
to somewhat smaller values \cite{footnote6}.

\begin{figure}[tb]
\begin{center}
\includegraphics[clip,width=7cm]{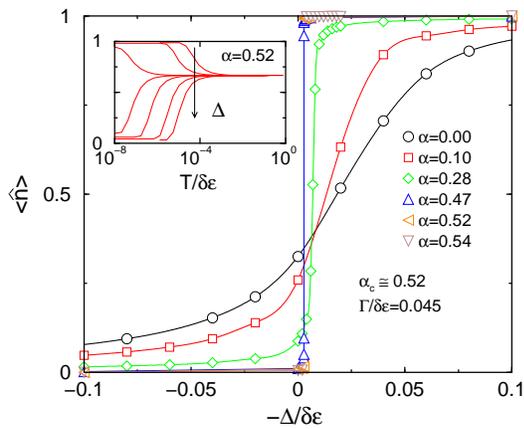}
\end{center}
\vskip0.1cm
\caption{\label{fig:steps}
Charging steps, computed using NRG for a relatively small
hybridization. For these parameters $T_K/\delta\epsilon \sim 10^{-10}$
at $\alpha\approx \alpha_c$.  Inset: Temperature dependence of the occupation
number $\langle\hat{n}\rangle$ at the critical dissipation,
$\alpha\approx\alpha_c$, for $\Delta/\delta\epsilon=-0.0028800$,
$-0.0028300$,
$-0.0028200$,
$-0.0028198$,
$-0.0028190$,
$-0.0028140$,
$-0.0028000$  (bottom to top).
}
\end{figure}

To obtain a more quantitative picture of the phase transition, we
performed numerical renormalization group calculations.
We used the Anderson Hamiltonian,
Eqs.~(\ref{eq:H_dot}) and (\ref{eq:T}), and represented the field
$\varphi$ by fermionic density fluctuations \cite{Pascal}. The
computed $T=0$ charging steps are shown in Fig.~\ref{fig:steps}. The
steps get sharper and sharper as we approach $\alpha_c$,
and finally a sudden step appears for $\alpha > \alpha_c$.

\begin{figure}[b]
\begin{center}
\includegraphics[clip,width=7cm]{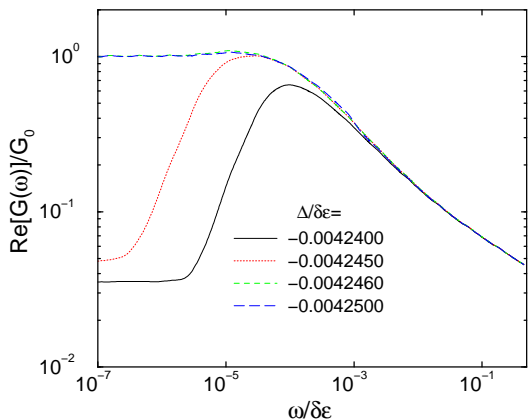}
\end{center}
\vskip0.1cm \caption{\label{ocond}
$T=0$ AC conductivity of the SET in the localized phase
for $\alpha = 0.75$ and $\Gamma / \delta\epsilon = 0.5$ ($G_0\sim G_Q$).
}
\end{figure}

The $T=0$ temperature AC conductivity is shown in Fig.~\ref{ocond}
in the charge localized phase. The AC conductivity clearly shows the
 Kondo resonance, whose width remains finite as one approaches the charge step from the
$\langle n\rangle \approx 1$ side, while a {\em dip} of width
$\sim |\Delta|$ develops on the $\langle n\rangle \approx 0$ side,
consistent with  the $G(\omega = 0)$ conductance having a jump. Similar non-monotonic behavior
should appear as a function of temperature.

{\gm
To achieve a large dissipation strength we propose to make a shallow
two-dimensional electron gas (2DEG) and fabricate a highly resistive
metallic  top gate electrode just above the dot, with a resistance
larger than $R_Q$. According to our estimates, dissipation strength
of the  order of $\alpha \approx 1$ can be reached in this way. The
SET can be then tuned through the quantum phase transition by either
continuously changing the tunneling $V$, or by depleting a second
2DEG positioned below the dot and thereby changing the total
capacitance of the dot and hence the value of $\alpha$.}

In summary, we have shown that sufficiently strong dissipation in
the gate electrodes can drive the SET through a quantum phase
transition into a state  where  charge degrees of freedom become
localized while spin fluctuations lead to a Kondo effect. In this
state both the conductance and the expectation value of the charge
on the SET display a jump at temperature $T=0$, while at higher
temperatures an anomalous scaling of the Coulomb blockade peaks is
predicted. We estimate that this quantum phase transition can be
detected by coupling a highly resistive gate electrode to a SET in a
shallow 2DEG.

We are grateful to P. Simon, K. Le Hur, Q. Si,  O. Sauret, A. Zaikin and 
Y. Nazarov  for valuable discussions.  This research has been supported
by NSF-MTA-OTKA Grant No. INT-0130446, Hungarian Grants
Nos. 
T046303, 
NF061726,  D048665,  and  T048782,
the European 'Spintronics' RTN HPRN-CT-2002-00302,
 and at Stanford University by NSF CAREER Award DMR-0349354
and a Packard Fellowship. L.B. is a grantee of the J\'anos Bolyai
Scholarship.


\vspace{-0.2cm}
\begin{references}
\vspace{-.2cm}
\bibitem{Metal_SET_reviews} 
For early reviews, see  e.g. 
G. Sch\"on and A.D. Zaikin, Phys. Rep. {\bf 198}, 237 (1990), 
or G.-L. Ingold and Y.V. Nazarov, in: Single Charge Tunneling, ed. 
by H. Grabert and M. Devoret, NATO ASI Series B, vol. 294, pp. 21-107 (Plenum, 1992). 
\bibitem{Metal_SET}
See also: S. V. Panyukov and A. D. Zaikin, Phys. Rev. Lett. {\bf 67}, 3168 (1991);
G. Falci, G. Sch\"on, and G. T. Zimanyi, Phys. Rev. Lett. {\bf 74}, 3257 (1995);
M. Kindermann and Yu. V. Nazarov, Phys. Rev. Lett. {\bf 91}, 136802 (2003).
\bibitem{Metal_SET_experiments}
P. Joyez et al., 
Phys. Rev. Lett. {\bf 79}, 1349-1352 (1997); 
D. Chouvaev et al.,  Phys. Rev. B {\bf 59}, 10599 (1999);
C. Wallisser et al., Phys. Rev. B {\bf 66}, 125314 (2002).
\bibitem{Matveev}
K.A. Matveev,
Phys. Rev. B {\bf 51}, 1743 (1995).
\bibitem{Costi} 
T. A. Costi, Phys. Rev. B {\bf 64}, 241310(R) (2001).
\bibitem{LeHur}
K. Le Hur, Phys. Rev. Lett. {\bf 92}, 196804 (2004).
\bibitem{Pascal}
L. Borda, G. Zarand, and P. Simon, Phys. Rev. B {\bf 72}, 155311 (2005);
M.-R. Li, K. Le Hur, and W. Hofstetter, Phys. Rev. Lett. {\bf 95}, 
086406 (2005).
\bibitem{Karyn} It has been argued earlier
based on calculations for a simple spinless model that
the dissipative transition should survive even in this limit:
K. Le Hur and M.-R. Li Phys. Rev. B {\bf 72}, 073305 (2005).
\bibitem{footnote1}
The analysis would be very similar for the transition between a singly- and 
doubly-occupied state.
\bibitem{footnote2}
With this normalization $V\sim \varrho_0^{1/2}$, with $\varrho_0$ the
density of states in the leads. 
\bibitem{footnote3} We shall neglect dissipation  on
source and drain electrodes which we assume not to be highly resistive.
\bibitem{footnote4} 
The constant  2/3 appears naturally along the calculations
and is related to the (classical) expectation value of the charge at the degeneracy point.
\bibitem{PofE}
M.H. Devoret {\em et al.}, Phys. Rev. Lett. {\bf 64}, 1824 (1990);
S. M. Girvin, {\em et al.},
Phys. Rev. Lett. {\bf 64}, 3183-3186 (1990).
\bibitem{Delft} J. von Delft and H. Schoeller, Annalen Phys. {\bf 7}, 225 (1998).
\bibitem{Cardy} J. Cardy, Scaling and Renormalization in Statistical
Physics (Cambridge University Press, Cambridge, 1996).
\bibitem{Kotliar} Q. Si and G. Kotliar, Phys. Rev. Lett. {\bf 70}, 3143 (1993).
\bibitem{footnote5} The scaling equation for $\tilde\Delta$
in  Ref.~\onlinecite{Kotliar} breaks SU(2) invariance, and some care is needed.
\bibitem{unpub} G. Zar\'and {\em et al.}, unpublished.
\bibitem{Kosterlitz} J.M. Kosterlitz, J. Phys. C {\bf 7}, 1046 (1974).
\bibitem{Wilson} H.R. Krishna-murthy {\em et al.}, Phys. Rev. B {\bf
  21}, 1003 (1980).
\bibitem{Kotliar2} G. Kotliar and Q. Si, Phys. Rev. B {\bf 53}, 12373 (1996).
\bibitem{Nozieres} P. Nozi\`eres, J. Low Temp. Phys. {\bf 17}, 31 (1974).
\bibitem{footnote6} Slightly below the critical
coupling $\alpha_c^{(0)}$, the tunneling $v$ is suppressed so much
that the Kondo scale becomes larger than the mixed valence scale
$T^{**}$. At the Kondo fixed point, however, the tunneling becomes
marginal, and smaller values of $\alpha$ are sufficient to localize
charge fluctuations.  Note  that  Eq.~(\ref{eq:v}) is inappropriate
in this  Kondo regime, and a strong coupling analysis is needed to
obtain the above picture.
\end{references}
\end{document}